\documentclass[a4paper, oneside, twocolumn, notitlepage, 10pt]{extarticle_ecoc}
\usepackage{ecoc}
\usepackage[helvet]{sfmath}

\newcommand{\SetCapsType}{normalcaps}

\usepackage[dvipsnames]{xcolor}
\usepackage{float}
\AtEndPreamble{\DeclareSymbolFont{operators}{OT1}{\rmdefault}{m}{n}}
\usepackage[per-mode=symbol]{siunitx}
\usepackage{lipsum}
\usepackage{stfloats}
\usepackage{amsmath}

\usepackage{glossaries}
\glsdisablehyper

\makeatletter
\providecommand{\SetCapsType}{smallcaps}

\long\def\@scTrue{smallcaps}
\long\def\@scFalse{normalcaps}
\newcommand{\acroSCaps}[1]{%
 \begingroup
  \ifx\SetCapsType\@scTrue 
    \textsc{#1}%
  \else
    \MakeUppercase{#1}%
  \fi
  \endgroup
}
\makeatother

\newcommand{\nAcronym}[4][]{%
	\newacronym[#1]{#2}{#3}{#4}
}

\makeatletter
\@ifpackageloaded{babel}{%
    \newcommand{\usuk}[2]{%
        \iflanguage{USenglish}{#1}{#2}%
    }%
}{%
    \newcommand{\usuk}[2]{%
        #1%
    }%
}%
\makeatother

\usepackage[shortcuts]{extdash}
\nAcronym{2A8PSK}{\acroSCaps{2a8psk}}{2-ary amplitude 8-ary phaseshift keying}

\nAcronym{3CCMCF}{\acroSCaps{3cc-mcf}}{3-core coupled-core multi-core fiber}

\nAcronym{4D}{\acroSCaps{4d}}{four-dimensional}
\nAcronym{4D64PRS}{\acroSCaps{4d-64prs}}{\usuk{four-dimensional 64-ary polarization-ring-switching}{four-dimensional 64-ary polarisation-ring-switching}}

\nAcronym{5B4D2A8PSK}{\acroSCaps{5b4d-2a8psk}}{5-bit four-dimensional two-amplitude 8-ary phase-shift keying}

\nAcronym{6B4D2A8PSK}{\acroSCaps{6b4d-2a8psk}}{6-bit four-dimensional two-amplitude 8-ary phase-shift keying}

\nAcronym{8D}{\acroSCaps{8d}}{eight-dimensional}\nAcronym{8D2048PRS}{\acroSCaps{8d-2048prs}}{eight-dimensional 2048-ary polarization-ring-switching}
\nAcronym{8D2048PRST1}{\acroSCaps{8d-2048prs-t1}}{eight-dimensional 2048-ary polarization-ring-switching type 1}
\nAcronym{8D2048PRST2}{\acroSCaps{8d-2048prs-t2}}{eight-dimensional 2048-ary polarization-ring-switching type 2}

\nAcronym{ABC}{\acroSCaps{abc}}{automatic bias controller}
\nAcronym{ADC}{\acroSCaps{adc}}{analog-to-digital converter}
\nAcronym{AIR}{\acroSCaps{air}}{achievable information rate}
\nAcronym{AOM}{\acroSCaps{aom}}{acoustic optical modulator}
\nAcronym{API}{\acroSCaps{api}}{application programming interface}
\nAcronym{AR}{\acroSCaps{ar}}{achievable rate}
\nAcronym{ASE}{\acroSCaps{ase}}{amplified spontaneous emission}
\nAcronym{ASK}{\acroSCaps{ask}}{amplitude-shift keying}
\nAcronym{ASIC}{\acroSCaps{asic}}{application-specific integrated circuit}
\nAcronym{AWGN}{\acroSCaps{awgn}}{additive white Gaussian noise}

\nAcronym{BCH}{\acroSCaps{bch}}{Bose-Chaudhuri-Hocquenghem}
\nAcronym{BER}{\acroSCaps{ber}}{bit error rate}
\nAcronym{BICM}{\acroSCaps{bicm}}{bit-interleaved coded modulation}
\nAcronym{BMD}{\acroSCaps{bmd}}{bit-metric decoding}
\nAcronym{BPD}{\acroSCaps{bpd}}{balanced photo-diode}
\nAcronym{BPS}{\acroSCaps{bps}}{blind phase search}
\nAcronym{BPSK}{\acroSCaps{bpsk}}{binary phase-shift keying}
\nAcronym{BRGC}{\acroSCaps{brgc}}{binary reflected Gray code}
\nAcronym{BTB}{\acroSCaps{btb}}{back-to-back}

\nAcronym{CAGR}{\acroSCaps{cagr}}{compound annual growth rate}
\nAcronym{CCDM}{\acroSCaps{ccdm}}{constant composition distribution matching}
\nAcronym{CCF}{\acroSCaps{ccf}}{\usuk{coupled-core fiber}{coupled-core fibre}}
\nAcronym{CD}{\acroSCaps{cd}}{chromatic dispersion}
\nAcronym{CIR}{\acroSCaps{cir}}{channel impulse response}
\nAcronym{CMA}{\acroSCaps{cma}}{constant modulus algorithm}
\nAcronym{CMUX}{\acroSCaps{cmux}}{core multiplexer}
\nAcronym{ChUT}{\acroSCaps{chut}}{channel under test}
\nAcronym{CUT}{\acroSCaps{cut}}{channel under test}
\nAcronym{CPE}{\acroSCaps{cpe}}{carrier phase estimation}
\nAcronym{CPU}{\acroSCaps{cpu}}{central processing unit}
\nAcronym{CW}{\acroSCaps{cw}}{continuous wave}

\nAcronym{DA}{\acroSCaps{da}}{driver amplifier}
\nAcronym{DAC}{\acroSCaps{dac}}{digital-to-analog converter}
\nAcronym{DCF}{\acroSCaps{dcf}}{\usuk{dispersion compensated fiber}{dispersion compensated fibre}}
\nAcronym{DCI}{\acroSCaps{dci}}{data center interconnect}
\nAcronym{DDLMS}{\acroSCaps{dd-lms}}{decision-directed least mean square}
\nAcronym{DFB}{\acroSCaps{dfb}}{distributed feedback}
\nAcronym{DGD}{\acroSCaps{dgd}}{differential group delay}
\nAcronym{DH}{\acroSCaps{dh}}{digital holography}
\nAcronym{DM}{\acroSCaps{dm}}{distribution matcher}
\nAcronym{DMA}{\acroSCaps{dma}}{direct memory access}
\nAcronym{DMD}{\acroSCaps{dmd}}{differential mode delay}
\nAcronym{DMG}{\acroSCaps{dmg}}{differential modal gain}
\nAcronym{DMGD}{\acroSCaps{dmgd}}{differential mode group delay}
\nAcronym{DML}{\acroSCaps{dml}}{directly-modulated laser}
\nAcronym{DP}{\acroSCaps{dp}}{\usuk{dual-polarization}{dual-polarisation}}
\nAcronym{DPC}{\acroSCaps{dpc}}{digital pre-compensation}
\nAcronym{DPE}{\acroSCaps{dpe}}{digital pre-emphasis}
\nAcronym{DPIQ}{\acroSCaps{dp-iqm}}{\usuk{dual-polarization IQ-modulator}{dual-polarisation IQ-modulator}}
\nAcronym{DRA}{\acroSCaps{dra}}{distributed Raman amplifier}
\nAcronym{DRE}{\acroSCaps{dre}}{digital resolution enhancer}
\nAcronym{DSF}{\acroSCaps{dsf}}{\usuk{dispersion-shifted fiber}{dispersion-shifted fibre}}
\nAcronym{DSO}{\acroSCaps{dso}}{digital sampling oscilloscope}
\nAcronym{DSP}{\acroSCaps{dsp}}{digital signal processing}
\nAcronym{DPLL}{\acroSCaps{dpll}}{digital phase-locked loop}
\nAcronym{DUT}{\acroSCaps{dut}}{device-under-test}
\nAcronym{DWDM}{\acroSCaps{dwdm}}{dense wavelength-division multiplexing}

\nAcronym{EAM}{\acroSCaps{eam}}{electro-absorption modulator}
\nAcronym{ECL}{\acroSCaps{ecl}}{external cavity laser}
\nAcronym{ED}{\acroSCaps{ed}}{Eucledian distance}
\nAcronym{EDFA}{\acroSCaps{edfa}}{\usuk{erbium-doped fiber amplifier}{erbium-doped fibre amplifier}}
\nAcronym{ENOB}{\acroSCaps{enob}}{effective number of bits}
\nAcronym{ESS}{\acroSCaps{ess}}{enumerative sphere shaping}

\nAcronym{FD}{\acroSCaps{fd}}{frequency domain}
\nAcronym{FDE}{\acroSCaps{fde}}{\usuk{frequency domain equalizer}{frequency domain equaliser}}
\nAcronym{FEC}{\acroSCaps{fec}}{forward error correction}
\nAcronym{FFT}{\acroSCaps{fft}}{fast Fourier transform}
\nAcronym{FIR}{\acroSCaps{fir}}{finite impulse response}
\nAcronym{FMF}{\acroSCaps{fmf}}{\usuk{few-mode fiber}{few-mode fibre}}
\nAcronym[plural=FM-MCF, firstplural=\usuk{few-mode multi-core fibers}{few-mode multi-core fibres}]{FM-MCF}{\acroSCaps{fm-mcf}}{\usuk{few-mode multi-core fiber}{few-mode multi-core fibre}}
\nAcronym{FPGA}{\acroSCaps{fpga}}{field-programmable gate array}
\nAcronym{FWM}{\acroSCaps{fwm}}{four-wave mixing}

\nAcronym{GD}{\acroSCaps{gd}}{group delay}
\nAcronym{GI}{\acroSCaps{gi}}{graded-index}
\nAcronym{GFF}{\acroSCaps{gff}}{gain flattening filter}
\nAcronym{GIMMF}{\acroSCaps{gi-mmf}}{\usuk{graded-index multi-mode fiber}{graded-index multi-mode fibre}}
\nAcronym{GMI}{\acroSCaps{gmi}}{\usuk{generalized mutual information}{generalised mutual information}}
\nAcronym{GNSE}{\acroSCaps{gnse}}{generalized nonlinear Schr\"{o}dinger equation}
\nAcronym{GPU}{\acroSCaps{gpu}}{graphics processing unit}
\nAcronym{GS}{\acroSCaps{gs}}{geometric shaping}
\nAcronym{GV}{\acroSCaps{gv}}{group-velocity}
\nAcronym{GVD}{\acroSCaps{gvd}}{group-velocity dispersion}
\nAcronym{GPIO}{\acroSCaps{gpio}}{general purpose input output}

\nAcronym{HDFEC}{\acroSCaps{hd-fec}}{hard decision forward error correction}

\nAcronym[plural=IL, firstplural=insertion losses (\textsc{il})]{IL}{\acroSCaps{il}}{insertion loss}
\nAcronym{IFFT}{\acroSCaps{ifft}}{inverse fast Fourier transform}
\nAcronym{IMDD}{\acroSCaps{im/dd}}{intensity-modulation direct-detection}
\nAcronym{ISI}{\acroSCaps{isi}}{intersymbol interference}

\nAcronym{KK}{\acroSCaps{kk}}{Kramers-Kronig}

\nAcronym{LCOS}{\acroSCaps{LCoS}}{liquid crystal on silicon}
\nAcronym{LDPC}{\acroSCaps{ldpc}}{low-density parity-check}
\nAcronym{LEAF}{\acroSCaps{leaf}}{\usuk{large effective area fiber}{large effective area fibre}}
\nAcronym{LFSR}{\acroSCaps{lfsr}}{linear-feedback shift register}
\nAcronym{LMS}{\acroSCaps{lmf}}{least means square}
\nAcronym{LLR}{\acroSCaps{llr}}{log-likelihood ratio}
\nAcronym{LO}{\acroSCaps{lo}}{local oscillator}
\nAcronym{LP}{\acroSCaps{lp}}{\usuk{linearly polarized}{linearly polarised}}
\nAcronym{LSPS}{\acroSCaps{lsps}}{\usuk{loop-synchronized polarization scrambler}{loop-synchronised polarisation scrambler}}
\nAcronym{LUT}{\acroSCaps{lut}}{lookup table}

\nAcronym{MB}{\acroSCaps{mb}}{Maxwell-Bolzmann}
\nAcronym{MCF}{\acroSCaps{mcf}}{\usuk{multi-core fiber}{multi-core fibre}}
\nAcronym{MDG}{\acroSCaps{mdg}}{mode dependent gain}
\nAcronym[plural=MDL, firstplural=mode-dependent losses (\textsc{mdl})]{MDL}{\acroSCaps{mdl}}{mode-dependent loss}
\nAcronym{MDM}{\acroSCaps{mdm}}{mode division multiplexing}
\nAcronym{MEMS}{\acroSCaps{mems}}{micro-electro-mechanical systems}
\nAcronym{MF}{\acroSCaps{mf}}{matched filter}
\nAcronym{MI}{\acroSCaps{mi}}{mutual information}
\nAcronym{MIMO}{\acroSCaps{mimo}}{multiple-input multiple-output}
\nAcronym{MMA}{\acroSCaps{mma}}{multi-modulus algorithm}
\nAcronym{MMF}{\acroSCaps{mmf}}{\usuk{multi-mode fiber}{multi-mode fibre}}
\nAcronym{MMSE}{\acroSCaps{mmse}}{minimum mean squared error}
\nAcronym{MPLC}{\acroSCaps{mplc}}{multi-plane light converter}
\nAcronym{MSE}{\acroSCaps{mse}}{mean squared error}
\nAcronym{MUX}{\acroSCaps{mux}}{multiplexer}
\nAcronym{MZM}{\acroSCaps{mzm}}{Mach-Zehnder modulator}
\nAcronym{MZI}{\acroSCaps{mzi}}{Mach-Zehnder interferometer}

\nAcronym{NA}{\acroSCaps{na}}{numerical aperture}
\nAcronym{NF}{\acroSCaps{nf}}{noise figure}
\nAcronym{NGMI}{\acroSCaps{ngmi}}{\usuk{normalized generalized mutual information}{normalised generalised mutual information}}
\nAcronym{NLSE}{\acroSCaps{nlse}}{nonlinear Schr\"{o}ding equation}
\nAcronym{NIR}{\acroSCaps{nir}}{near-infrared}

\nAcronym{OBTB}{\acroSCaps{obtb}}{optical back-to-back}
\nAcronym{ODE}{\acroSCaps{ode}}{ordinary differential equation}
\nAcronym{ODL}{\acroSCaps{odl}}{optical delay line}
\nAcronym{OFDR}{\acroSCaps{ofdr}}{optical frequency-domain reflectometer}
\nAcronym{OFDM}{\acroSCaps{ofdm}}{orthogonal frequency division multiplexing}
\nAcronym{OMFT}{\acroSCaps{omft}}{optical multi-format transmitter}
\nAcronym{OOK}{\acroSCaps{ook}}{on-off keying}
\nAcronym{OPLL}{\acroSCaps{opll}}{optical phase-locked loop}
\nAcronym{OSA}{\acroSCaps{osa}}{\usuk{optical spectrum analyzer}{optical spectrum analyser}}
\nAcronym{OSNR}{\acroSCaps{osnr}}{optical signal-to-noise ratio}
\nAcronym{OTDR}{\acroSCaps{otdr}}{optical time-domain reflectometer}
\nAcronym{OTF}{\acroSCaps{otf}}{optical tunable filter}
\nAcronym{OVNA}{\acroSCaps{ovna}}{\usuk{optical vector network analyzer}{optical vector network analyser}}

\nAcronym{PAM}{\acroSCaps{pam}}{pulse-amplitude modulation}
\nAcronym{PAS}{\acroSCaps{pas}}{probabilistic amplitude shaping}
\nAcronym{PAPR}{\acroSCaps{papr}}{peak-to-avarage power ratio}
\nAcronym{PBC}{\acroSCaps{pbc}}{\usuk{polarization beam combiner}{polarisation beam combiner}}
\nAcronym{PBS}{\acroSCaps{pbs}}{polarization beam splitter}
\nAcronym{PCVD}{\acroSCaps{pcvd}}{plasma chemical vapor depostion}
\nAcronym{PD}{\acroSCaps{pd}}{photodiode}
\nAcronym{PDL}{\acroSCaps{pdl}}{polarization-dependent loss}
\nAcronym{PDM}{\acroSCaps{pdm}}{\usuk{polarization-division multiplexing}{polarisation-division multiplexing}}
\nAcronym{PL}{\acroSCaps{pl}}{photonic lantern}
\nAcronym{PMBPSK}{\acroSCaps{pm-bpsk}}{polarization-multiplexed binary phase-shift-keying}
\nAcronym{PMQPSK}{\acroSCaps{pm-qpsk}}{polarization-multiplexed quaternary phase-shift-keying}
\nAcronym{PM8QAM}{\acroSCaps{pm-8qam}}{polarization-multiplexed 8-ary quadrature amplitude modulation}
\nAcronym{PMD}{\acroSCaps{pmd}}{polarization mode dispersion}
\nAcronym{PNOB}{\acroSCaps{pnob}}{physical number of bits}
\nAcronym{PON}{\acroSCaps{pon}}{passive-optical network}
\nAcronym{PRBS}{\acroSCaps{prbs}}{pseudorandom bit sequence}
\nAcronym{PS}{\acroSCaps{ps}}{probabilistic shaping}
\nAcronym{PSK}{\acroSCaps{psk}}{phase-shift keying}
\nAcronym{PSP}{\acroSCaps{psp}}{\usuk{principal states of polarization}{principal states of polarisation}}

\nAcronym{QAM}{\acroSCaps{qam}}{quadrature amplitude modulation}
\nAcronym{QPSK}{\acroSCaps{qpsk}}{quaternary phase-shift-keying}

\nAcronym{RAM}{\acroSCaps{ram}}{random-access memory}
\nAcronym{RF}{\acroSCaps{rf}}{radio frequency}
\nAcronym{RRC}{\acroSCaps{rrc}}{root-raised-cosine}
\nAcronym{ROADM}{\acroSCaps{roadm}}{reconfigurable optical add-drop multiplexer}
\nAcronym{ROI}{\acroSCaps{roi}}{region of interest}

\nAcronym{SA}{\acroSCaps{sa}}{simulated annealing}
\nAcronym{SamPerSym}{\acroSCaps{sps}}{samples per symbol}
\nAcronym{SBS}{\acroSCaps{sbs}}{stimulated Brillouin scattering}
\nAcronym{SDFEC}{\acroSCaps{sd-fec}}{soft decision forward error correction}
\nAcronym{SDM}{\acroSCaps{sdm}}{space-division multiplexing}
\nAcronym{SE}{\acroSCaps{se}}{spectral efficiency}
\nAcronym{SER}{\acroSCaps{ser}}{symbol error rate}
\nAcronym{SI}{\acroSCaps{si}}{step index}
\nAcronym{SLM}{\acroSCaps{slm}}{spatial light modulator}
\nAcronym{SMF}{\acroSCaps{smf}}{\usuk{single-mode fiber}{single-mode fibre}}
\nAcronym{SNR}{\acroSCaps{snr}}{signal-to-noise ratio}
\nAcronym{SOA}{\acroSCaps{soa}}{semiconductor optical amplifier}
\nAcronym[firstplural=\usuk{states of polarization (\acroSCaps{sop})}{states of polarisation (\acroSCaps{sop})}]{SOP}{\acroSCaps{sop}}{\usuk{state of polarization}{state of polarisation}}
\nAcronym{SPM}{\acroSCaps{spm}}{self-phase modulation}
\nAcronym{SPS}{\acroSCaps{sps}}{\usuk{synchronized polarization scrambler}{synchronised polarisation scrambler}}
\nAcronym{SRS}{\acroSCaps{srs}}{stimulated Raman scattering}
\nAcronym{SSFM}{\acroSCaps{ssfm}}{split-step Fourier method}
\nAcronym{SSMF}{\acroSCaps{ssmf}}{\usuk{standard single-mode fiber}{standard single-mode fibre}}
\nAcronym{STL}{\acroSCaps{stl}}{swept tunable laser}
\nAcronym{SVD}{\acroSCaps{svd}}{singular value decomposition}
\nAcronym{SWI}{\acroSCaps{swi}}{swept wavelength interferometry}

\nAcronym{TD}{\acroSCaps{td}}{time domain}
\nAcronym{TDE}{\acroSCaps{tde}}{\usuk{time domain equalizer}{time domain equaliser}}
\nAcronym{TDMSDM}{\acroSCaps{tdm-sdm}}{time-domain multiplexed space-division multiplexing}
\nAcronym{TE}{\acroSCaps{te}}{transverse electric}
\nAcronym{TIA}{\acroSCaps{tia}}{trans-impedance amplifier}
\nAcronym{TM}{\acroSCaps{TM}}{transverse magnetic}
\nAcronym{TH4D}{\acroSCaps{th-4d}}{time domain hybrid four-dimensional}
\nAcronym{TH4D2A8PSK}{\acroSCaps{th-4d-2a8psk}}{time domain hybrid four-dimensional two-amplitude eight-phase shift keying}

\nAcronym{VCSEL}{\acroSCaps{vcsel}}{vertical-cavity surface emitting laser}
\nAcronym{VOA}{\acroSCaps{voa}}{variable optical attenuator}

\nAcronym{WDM}{\acroSCaps{wdm}}{wavelength-division multiplexing}
\nAcronym{WGA}{\acroSCaps{wga}}{weakly guiding approximation}
\nAcronym{WGN}{\acroSCaps{wgn}}{white Gaussian noise}
\nAcronym{WSS}{\acroSCaps{wss}}{wavelength selective switch}

\nAcronym{XPM}{\acroSCaps{xpm}}{cross-phase modulation}
\nAcronym{XT}{\acroSCaps{xt}}{cross-talk}

\nAcronym{3DWG}{\acroSCaps{3dwg}}{3D-waveguide} 
\usepackage[capitalise]{cleveref}

\begin{document}
\selectlanguage{english}    


\title{Exploiting Angular Multiplexing for Polarization-diversity in Off-axis Digital Holography\vspace{-3mm}}%


\author{Sjoerd~van~der~Heide\textsuperscript{(1, *)},
        Rutger~van~Anrooij\textsuperscript{(1)},
        Menno~van~den~Hout\textsuperscript{(1)},
        Nicolas K. Fontaine\textsuperscript{(2)},\\
        Roland Ryf\textsuperscript{(2)},
        Haoshuo Chen\textsuperscript{(2)},
        Mikael Mazur\textsuperscript{(2)},
        Jose~Enrique~Antonio-L\'opez\textsuperscript{(3)},\\
        Juan~Carlos~Alvarado-Zacarias\textsuperscript{(3)}, 
        Ton~Koonen\textsuperscript{(1)},
        Rodrigo~Amezcua-Correa\textsuperscript{(3)},
        and Chigo~Okonkwo\textsuperscript{(1)}
}

\maketitle                  


\begin{strip}
 \begin{author_descr}

    \vspace{-2mm}\textsuperscript{(1)} High Capacity Optical Transmission Laboratory, Electro-Optical Communications Group,\\
    Eindhoven University of Technology, the Netherlands, \textsuperscript{(*)} s.p.v.d.heide@tue.nl
\\\textsuperscript{(2)} Nokia Bell Labs, 791 Holmdel Rd, Holmdel, NJ, 07733, USA
\\\textsuperscript{(3)} CREOL, The College of Optics and Photonics, University of Central Florida, USA
\vspace{-2mm}
 \end{author_descr}
\end{strip}
\setstretch{1.1}%
%
%
\begin{strip}
  \begin{ecoc_abstract}
    Digital holography measures the complex optical field and transfer matrix of a device, polarization-diversity is often achieved through spatial multiplexing. We introduce angular multiplexing, to increase flexibility in the optical setup. Comparatively, similar values for cross-talk and mode-dependent loss are measured for a photonic lantern.\vspace{-2mm}
  \end{ecoc_abstract}
\end{strip}
%
%
\section{Introduction}
\begin{figure}[b]
\vspace{-2mm}
\centering
\includegraphics[width=\columnwidth]{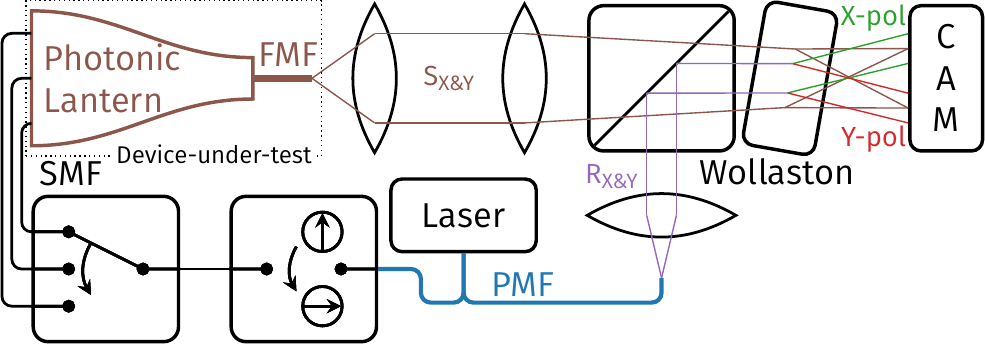}
\caption{Spatial multiplexing DH optical setup. Signal and reference light are combined, and a Wollaston prism is used to split polarizations across a near-infrared camera. Optical switches excite all inputs of the photonic lantern under test.}
\label{fig:setupSLO}
\vspace{-3mm}
\end{figure}

\Gls{SDM} has the potential to extend the capacity of optical fibers greatly beyond what is possible with current single-mode technologies. \Gls{SDM} uses spatially-diverse fibers such as few-mode, multi-mode, multi-core, coupled-core, and few-mode multi-core fibers, for which multiplexer devices to couple in to these fibers are required. For the development of these multiplexers such as \glspl{PL}\cite{velazquez2018scaling}, proper characterization tools for important metrics such as \gls{XT} and \gls{MDL} are very desirable.

Characterization of \gls{SDM} systems and devices can be done by methods such as analysis of equalizer taps\cite{winzer2011mimo} through which system-level metrics can be obtained, whilst \glspl{OVNA}\cite{NickOVNA, SimonOVNA} are also able to extract some valuable device-level metrics. Off-axis \gls{DH}\cite{ploschner2015seeing}, however, provides access to the full complex optical field for all device inputs and has recently been used to measure a plethora of \gls{SDM} devices\cite{fontaine2019laguerre, MikaelHolography, cmuxSjoerd, edfaJC}. To access both polarizations of the optical field, \gls{DH} is performed separately on each polarization, either by using multiple \gls{NIR} cameras or by using a different part of the camera, essentially multiplexing polarizations spatially. Alternatively, in the imaging field of research\cite{colomb2002polarization, Yuan:11, Ochiai:13}, the same part of a camera was used to perform \gls{DH} on a signal using two orthogonally polarized reference beams which were separated in angle, essentially multiplexing polarizations angularly.

\begin{figure}[b]
\vspace{-2mm}
\centering
\includegraphics[width=\columnwidth]{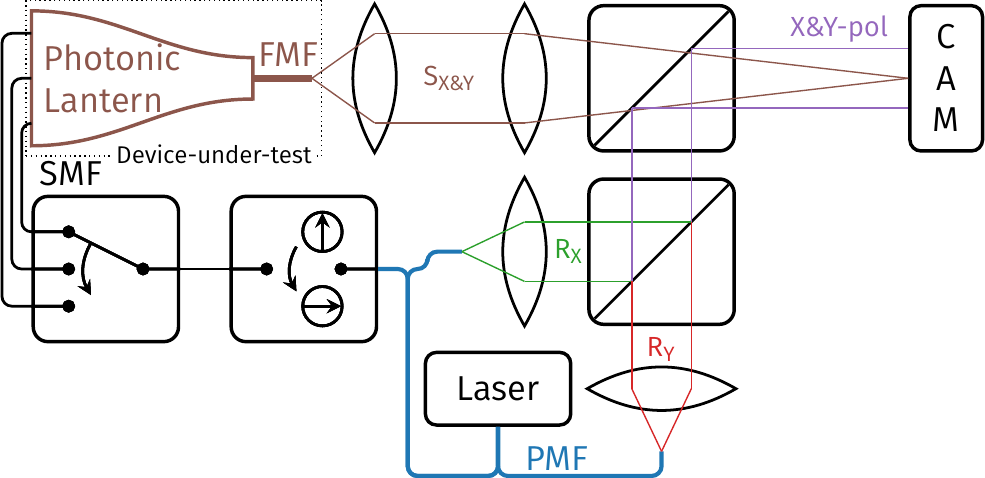}
\caption{Spatial multiplexing DH optical setup. Similar to \cref{fig:setupSLO}, but here two orthogonally-polarized reference beams are used instead of a Wollaston prism.}
\label{fig:setupDLO}
\vspace{-3mm}
\end{figure}

In this work, we introduce angular multiplexing to achieve polarization-diversity in \gls{DH} as an alternative to spatial multiplexing for measurements of the complex optical field and transfer matrix of an \gls{SDM} device. Both spatial and angular multiplexing schemes are discussed and compared through measurements of the same \gls{PL}. Similar \gls{XT} values of \SI{13.8}{dB} and \SI{14.0}{dB} are measured using spatial and angular multiplexing, respectively. Furthermore, we find very similar values of \SI{1.50}{dB} and \SI{1.45}{dB} for \gls{MDL}, confirming angular multiplexing is a viable alternative to spatial multiplexing. Angular multiplexing may lead to a simpler optical setup with fewer components in the signal path and greater magnification of the signal beam, increasing the resolution of measured optical fields.

\section{Polarization-diversity in Digital Holography}

\cref{fig:setupSLO} shows a commonly-used \gls{DH} setup where spatial multiplexing is used to achieve polarization-diversity. A 4f optical setup coherently combines \textit{signal} \SI{1550}{nm} laser light with \SI{100}{kHz} linewidth from a \gls{DUT}, in this case a \gls{PL}, with slightly angled (\textless 5 degrees) \textit{reference} light, producing a fringe pattern on a \gls{NIR} camera. Note that the lens setup is not drawn to scale to conserve space. A Wollaston prism with a 20 degree beam separation splits polarizations spatially across the camera, and optical switches are used to address all input ports and polarizations of the \gls{DUT}.

In contrast to the spatial multiplexing scheme of \cref{fig:setupSLO}, \cref{fig:setupDLO} shows the use of the entire camera surface for both polarizations through angular multiplexing. Instead of splitting light from both polarizations spatially, two orthogonally-polarized reference beams are used, each at a different slight angle with respect to the signal, making two different overlapping fringe patterns appear on the camera, which can be separated digitally. Fewer optical components are required in the signal path and potentially larger beam widths lead to greater flexibility.

\begin{figure}[t]
\vspace{-2mm}
\centering
\includegraphics[width=\columnwidth]{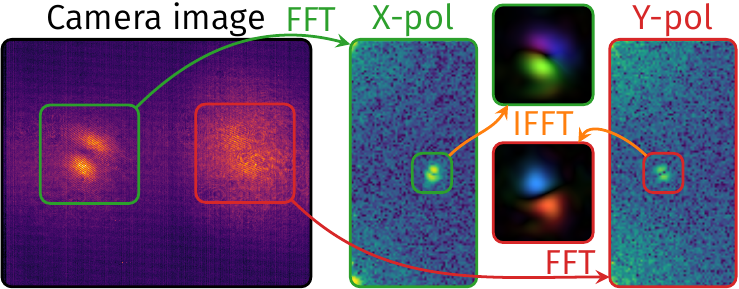}
\caption{Spatial multiplexing DH signal processing. Fringe patterns are distributed spatially across the camera and are cropped and Fourier transformed to the angular domain, revealing the holograms. Inverse Fourier transformation of the hologram gives the extracted complex fields.}
\label{fig:dspSLO}
\end{figure}

\section{Digital Holography Signal Processing}

The main elements of the \gls{DH} signal processing chain are depicted in \cref{fig:dspSLO} for measurements obtained using the spatial multiplexing setup detailed in \cref{fig:setupSLO}. Since the fringe patterns corresponding with the X- and Y-polarization are spatially separated on the \gls{NIR} camera, they can be cropped from frame and processed separately.
\vspace{-3mm}
\begin{equation}
|S+R|^2 = |S|^2 + |R|^2 + \mathbf{\color{ForestGreen}{SR^*}} + S^*R
\label{eq:DHSLO}
\vspace{-1mm}
\end{equation}
For each polarization, the intensity of the fringe pattern can be described by \cref{eq:DHSLO}. Both cropped images from the spatial domain are Fourier transformed to the angular domain. The detection of the signal and reference directly, $|S|^2$ and $|R|^2$, end up at DC in the angular domain since there is no angle between the signal or reference with respect to itself. The beating between signal and reference, the fringe pattern in the spatial domain, $SR^*$, however, manifests itself as a distinct part in the angular domain, with significant separation from DC-terms, which can be cropped in the angular domain to extract the desired term of \cref{eq:DHSLO}. It is inversely Fourier transformed, revealing the complex $SR^*$ term in the spatial domain, both in amplitude and phase, even though the camera only measures intensity. Given the reference is by approximation a plane wave, $SR^*$ equates to $S$, which, after processing both polarizations, results in the full polarization-diverse complex optical field for a given device excitation. Repeating the measurement for all possible input states through the optical switches in \cref{fig:setupSLO} provides the complete description for the device for a given wavelength.

Additional processing provides further insight. The optical fields can be \textit{digitally demultiplexed} to any desired modal basis using overlap-integrals, providing a transfer matrix between device input port and polarization and spatial mode of the desired modal basis. Important performance metrics such as \gls{XT} and \gls{MDL} can be obtained through analysis of this transfer matrix. Note that the desired modal basis can be any, for example a specific optical fiber of interest, essentially predicting how the device would behave when coupled or spliced with that specific fiber.
\vspace{-1mm}
\begin{equation}
\begin{aligned}
|S_{X\&Y}+R_X + R_Y|^2 = |S_{X\&Y}|^2 + |R_X|^2 + \\
|R_Y|^2 + \mathbf{\color{ForestGreen}{S_XR_X^*}} + S_X^*R_X + \mathbf{\color{ForestGreen}{S_YR_Y^*}} + S_Y^*R_Y 
\label{eq:DHDLO}
\end{aligned}
\vspace{-1mm}
\end{equation}
\begin{figure}[t]
\vspace{-2mm}
\centering
\includegraphics[width=\columnwidth]{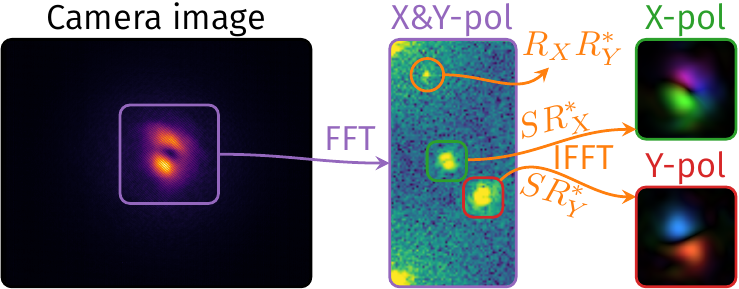}
\caption{Angular multiplexing DH signal processing. Fringe patterns overlay on the camera and are Fourier transformed together. Holograms for both polarizations appear at distinct locations in the angular domain, cropping and inverse Fourier transformation thereof gives the extracted complex fields.}
\label{fig:dspDLO}
\vspace{-1mm}
\end{figure}
The intensity of the camera frame in \cref{fig:dspDLO} for the angular multiplexing setup of \cref{fig:setupDLO} can be described by \cref{eq:DHDLO}. Cross-terms between X- and Y-polarizations are omitted since perfectly orthogonally-polarized beams do not interfere. In the spatial domain, both polarizations overlap, but after a \gls{FFT} they are separated in the angular domain since reference beams $R_X$ and $R_Y$ are placed under a slight angle, both with respect to the signal and each other. Since $S_XR_X^*$ and $S_YR_Y^*$ are separated in angular domain, similar to the signal processing chain for spatial multiplexing, they can be cropped and inversely Fourier transformed separately to reveal the complex optical field in spatial domain for both polarizations. The extracted fields are the same for spatial and angular multiplexing if the polarization axes of the two reference beams of the angular multiplexing setup are aligned to the Wollaston prism of the spatial multiplexing setup. Note that some interference between the orthogonally-polarized reference beams, $R_XR_Y^*$, is observed in the angular domain of \cref{fig:dspDLO}, indicating they were not perfectly orthogonal.

\section{Results and discussion}

\begin{figure}[t]
\centering
\includegraphics[width=\columnwidth]{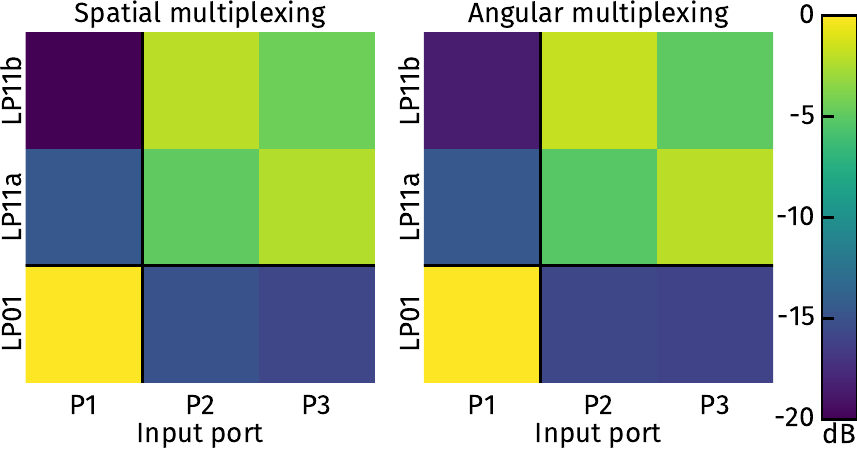}
\caption{Power transfer matrix from input port to output mode for a photonic lantern measured by both digital holography setups. Analysis reveals cross-talk values of \SI{-13.8}{dB} and \SI{-14.0}{dB}, and mode-dependent loss of \SI{1.50}{dB} and \SI{1.45}{dB}, for spatial and angular multiplexing, respectively.}
\label{fig:txplot}
\end{figure}

The same \gls{PL} was measured using both the spatial and angular multiplexing scheme for each of input ports and polarizations, and one complex optical field per output polarization is extracted using the signal processing techniques detailed in the previous section. Two of these output fields when port 3 polarization X was excited are displayed as an inset in \cref{fig:dspSLO} and \cref{fig:setupDLO} for spatial and angular multiplexing, respectively. The intensity of the image represents the amplitude of the optical field whilst a cyclic colormap is for the phase. It is observed that the extracted fields look very similar for both measurement setups. The extracted fields, 12 per measurement, look very similar for both multiplexing techniques, but are omitted in this paper to conserve space. This qualitatively confirms the validity of the angular multiplexing scheme by comparing it to the tried-and-tested spatial multiplexing setup.

To quantitatively assess the measurement results, the extracted fields are digitally demultiplexed to the Hermite-Gaussian \gls{LP} modal basis. The resulting 6x6 complex transfer matrix can be condensed to a 3x3 power transfer matrix, from input port to output \gls{LP}-mode, which is depicted in \cref{fig:txplot} for both spatial and angular multiplexing. Mode-group \gls{XT} is evaluated using this matrix and equates to \SI{-13.8}{dB} and \SI{-14.0}{dB} for spatial and angular multiplexing, respectively. \Gls{MDL}, calculated through singular value decomposition of the 6x6 complex transfer matrix, is \SI{1.50}{dB} and \SI{1.45}{dB} for spatial and angular multiplexing, respectively. Therefore, quantitatively, both measurement schemes reveal similar performance metrics for the same device, further validating the concept of angular multiplexing for polarization-diverse \gls{DH}.

These results are achieved using many of the same optical components for the spatial and angular multiplexing setup. However, in principle, the angular multiplexing scheme may benefit from using different components. For example, since no Wollaston prism is required, the reference beam width may be increased so it better resembles a plane wave. Also, since a larger area of the camera is available, greater magnification in the 4f lens setup can increase the number of pixels used for the signal and therefore increase resolution and dynamic range of the extracted optical fields. On the other hand, by incorporating two fringe patterns on the same part of the camera, dynamic range is reduced. 

Furthermore, one of the key attractions to \gls{DH} as a measurement technique is the low amount of optical components required to capture the signal light, keeping the influence of aberrations minimal. Thus it might prove beneficial that the angular multiplexing scheme does not have the Wollaston in the signal path, and only introduces a simple beam-splitter in the reference path.

Finally, some interference between both reference beams is observed in \cref{fig:dspDLO}, which could give meaningful insight in the reference beam amplitude and phase, potentially for correction of its deviation from a perfect plane-wave.


\section{Conclusion}
Angular multiplexing is introduced as a more flexible alternative to spatial multiplexing for polarization-diverse \acrlong{DH} measurements of \acrlong{SDM} devices. Both techniques are used to measure the same \acrlong{PL} to test the novel angular scheme against the tried-and-tested spatial multiplexing scheme. Similar values are measured for mode-group \acrlong{XT}, \SI{-13.8}{dB} versus \SI{-14.0}{dB}, and \acrlong{MDL}, \SI{1.50}{dB} versus \SI{1.45}{dB}, using spatial and angular multiplexing, respectively. Therefore, angular multiplexing for polarization-diverse \acrlong{DH} measurements is a valid alternative.

{\vspace{1mm}\footnotesize \setstretch{1.1} Partial funding from the NWO Gravitation Program on Research Center for Integrated Nanophotonics (Grant Number 024.002.033) and by TU/e-KPN Smart Two project.\par}

\newpage
\printbibliography
\vspace{-4mm}

\end{document}